\documentclass[twocolumn,amssymb,fleqn, twocolumns]{revtex4} %fleqn: make all eqns left alighed  ,showpacs
\usepackage{epsfig,amssymb,amsmath,graphicx,subfigure,hyperref  }  % pdfpages and graphicx are added   hyperref(set hyperlink for content)
\usepackage{color}

\usepackage{multirow} 
\usepackage{tabularx}

\begin{document}

\title{Dynamical effects of long-range interaction revealed in screened \\
Coulomb interacting ring systems}
\author{Zhenwei Yao}
\email{zyao@sjtu.edu.cn}
\affiliation{School of Physics and Astronomy, and Institute of Natural
Sciences, Shanghai Jiao Tong University, Shanghai 200240, China}
\begin{abstract}
Understanding the intriguing physical effects of long-range
interactions is a common theme in a host of physical systems. In this work, based
on the classical screened Coulomb interacting ring model, we investigate the
dynamical effects of the long-range interaction from the unique perspective of
analyzing the dynamical response of the system to disturbance. We reveal the
featured dynamics brought by the long-range interaction, including the
efficient transformation of the disturbance into the uniform global rotation of
the system, the suppression of the intrinsic noise, and the fast relaxation of
particle speed.
\end{abstract}

\maketitle

\section{Introduction}

Long-range interactions are widely seen in multiple fields, ranging from
astrophysics~\cite{lynden1968gravo, padmanabhan1990statistical,
joyce2011quasistationary}, hydrodynamics~\cite{lighthill1976flagellar,
chattopadhyay2009effect,dallaston2018discrete} to electrostatic and dipolar
systems~\cite{Walker2011, christodoulidi2014fermi,
juhasz2014random,jadhao2015coulomb, mauri2019thermal,yao2019command}. Notably,
electrostatic interaction
provides an important organizing force to fabricate exceedingly rich
morphologies in various soft matter systems~\cite{Holm2001, Levin2002,
messina2008electrostatics, xing2011poisson, yao2016electrostatics, gao2019electrostatic}. In long-range interacting systems, the strong coupling
among the elementary constituents brings in intriguing behaviors not found in
systems with short-range
interactions~\cite{bouchet2005classification,levin2005strange,
rocha2005entropy,pluchino2007nonergodicity,christodoulidi2014fermi}, and
meanwhile imposes a grand challenge in theoretical treatment of such
systems~\cite{joyce1971numerical, pakter2017entropy, cirto2018validity}.
Understanding the dynamical effects of long-range interaction is closely related
to a host of physical problems, such as the formation of long-range
order~\cite{toner1995long, zhang2013non, grzybowski2009self} and the
origin of nonlinear dynamical structures arising in various
contexts~\cite{Kadanoff1999,chen2012introduction,rutzel2003nonlinear,campa2014physics}.
While our knowledge about long-range interacting systems has been significantly
advanced by solving kinetic equations and mean-field models, the approach based
on precise numerical integration of the equations of motion has proven to be a
powerful tool to reveal the fundamental microscopic dynamics not
accessible by mean-field theories~\cite{boltzmann1964lectures, rapaport2004art,
feix2005universal, campa2014physics}.

In this work, we investigate the effect of long-range interaction on the
microscopic dynamics of interacting particles from the unique perspective of
analyzing the dynamical response of the system to disturbance imposed on a
single particle. Our model system consists of a collection of particles
confined on a ring that interact by the screened Coulomb potential; this
screened Coulomb interacting ring model is an invariant version of the
self-gravitating ring model and may be realized in colloidal
experiments~\cite{sota2001origin,sokolov2011hydrodynamic,nagar2014collective,williams2016transmission}. Adopting the screened Coulomb
potential, which can be experimentally realized in electrolyte
solutions~\cite{debye1923theory, Dobrynin2005}, allows us to conveniently control the range
of interaction, and thus highlight the unique effect of long-range interaction.
Restricting the configurational space of the particles on the periodic 1D
manifold excludes the subtle boundary effect. By analyzing the dynamical
evolution of the system upon the disturbance, we reveal the featured dynamics
driven by the long-range interaction, including the emergence of fine wave
structure over the trajectory of the disturbed particle and the proliferation of
dynamical modes in the energy spectrum.  Long-range interaction also leads to
efficient transformation of the disturbance into the uniform global rotation of
the system, suppression of the intrinsic noise, and fast relaxation of particle
speed.

\section{Model and Method}

\begin{figure}[t]  % h: put figure just in this position if possible
\centering 
\includegraphics[width=3.2in]{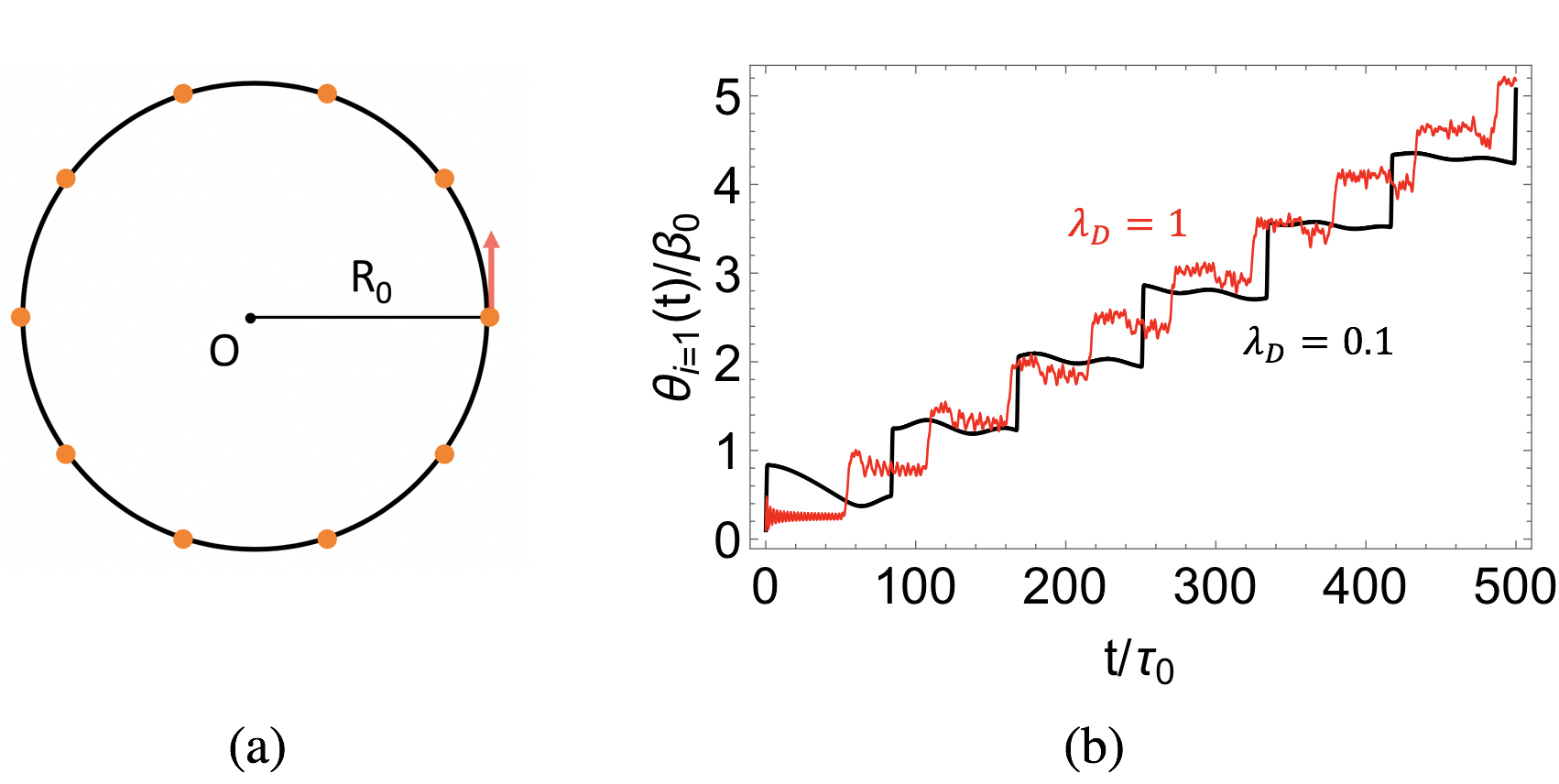}
  \caption{Trajectories of the disturbed particle in the ring model. (a) Schematic plot of the
  model consisting of identical point particles confined on the ring of radius
  $R_0$ that interact by the screened Coulomb potential. (b) The trajectories of
  the disturbed particle (labelled as $i=1$) for both cases of $\lambda_D=0.1$
  and $\lambda_D=1$. The initial speed is specified by
  $\Gamma=1$; see main text for more information. $N=100$. }
\label{schematic}
\end{figure}

The model system consists of $N$ identical point particles of mass $m$ confined
on a ring of radius $R_0$, as shown in fig.~\ref{schematic}(a). The particles
are initially evenly distributed on the ring with the lattice spacing $a_0=R_0
\beta_0$, where $\beta_0=2\pi/N$. The collection of geometrically
confined, interacting particles are treated as a pure
classical mechanical system, and we work in the microcanonical ensemble with
fixed number of particles and total energy. The microscopic dynamics of the
particles is governed by the Hamiltonian:
\begin{eqnarray}
  H = \sum_{i=1}^N \frac{\vec{L}_i^2}{2m R_0^2} + \frac{1}{2} \sum_{i\neq j}
  V(r_{ij}),\label{H}
\end{eqnarray}
where the screened Coulomb potential
\begin{eqnarray}
  V(r_{ij}) = V_0 \frac{e^{-r_{ij}/\lambda_D}}{r_{ij}}.
\end{eqnarray}
$\vec{L}_i$ is the angular momentum of the particle $i$, $r_{ij}$ is the
Euclidean distance between particles $i$ and $j$, and $\lambda_D$ is the
Debye screening
length~\cite{debye1923theory}. The range of interaction could be controlled by
the screening length $\lambda_D$. Dynamics is introduced by 
specifying an initial
velocity $\vec{v}_{{\textrm {ini}}}$ directly to a single particle; the remaining
particles are at rest. $\vec{v}_{{\textrm {ini}}}$ is tangent to the ring, and
its magnitude is $\Gamma v_0$, where $v_0$ is the characteristic speed.
$v_0=\sqrt{V_0/(m a_0)}$. The strength of the initial disturbance is determined
by the dimensionless quantity $\Gamma$. The disturbed particle serves as a probe
to reveal the dynamical effects of long-range interaction. To investigate the
relaxation process, we also discuss the case that all the particles are
disturbed.  The length, mass, and time are measured in the units of the lattice
spacing $a_0$, particle mass $m$ and $\tau_0$.  $\tau_0=a_0/v_0$.

We numerically integrate the equations of motion derived from the Hamiltonian in
eq.(\ref{H}) for long-time (up to hundreds of times of the characteristic time
scale) trajectories of motion at high precision. The time step $h$ is set to be
sufficiently fine to ensure that during a hundred million simulation steps the
variations of the total energy and the angular momentum are within one
thousandth with respect to the mean kinetic energy and the mean angular
momentum. Typical, $h=10^{-5}\tau_0$. No cut-off length is introduced in
dealing with the long-range interaction between particles. We investigate the
dynamical and statistical effects of long-range interaction by analyzing the
energy spectra and some key dynamical parameters. 

\begin{figure}[t]  % h: put figure just in this position if possible
\centering 
\includegraphics[width=3.4in]{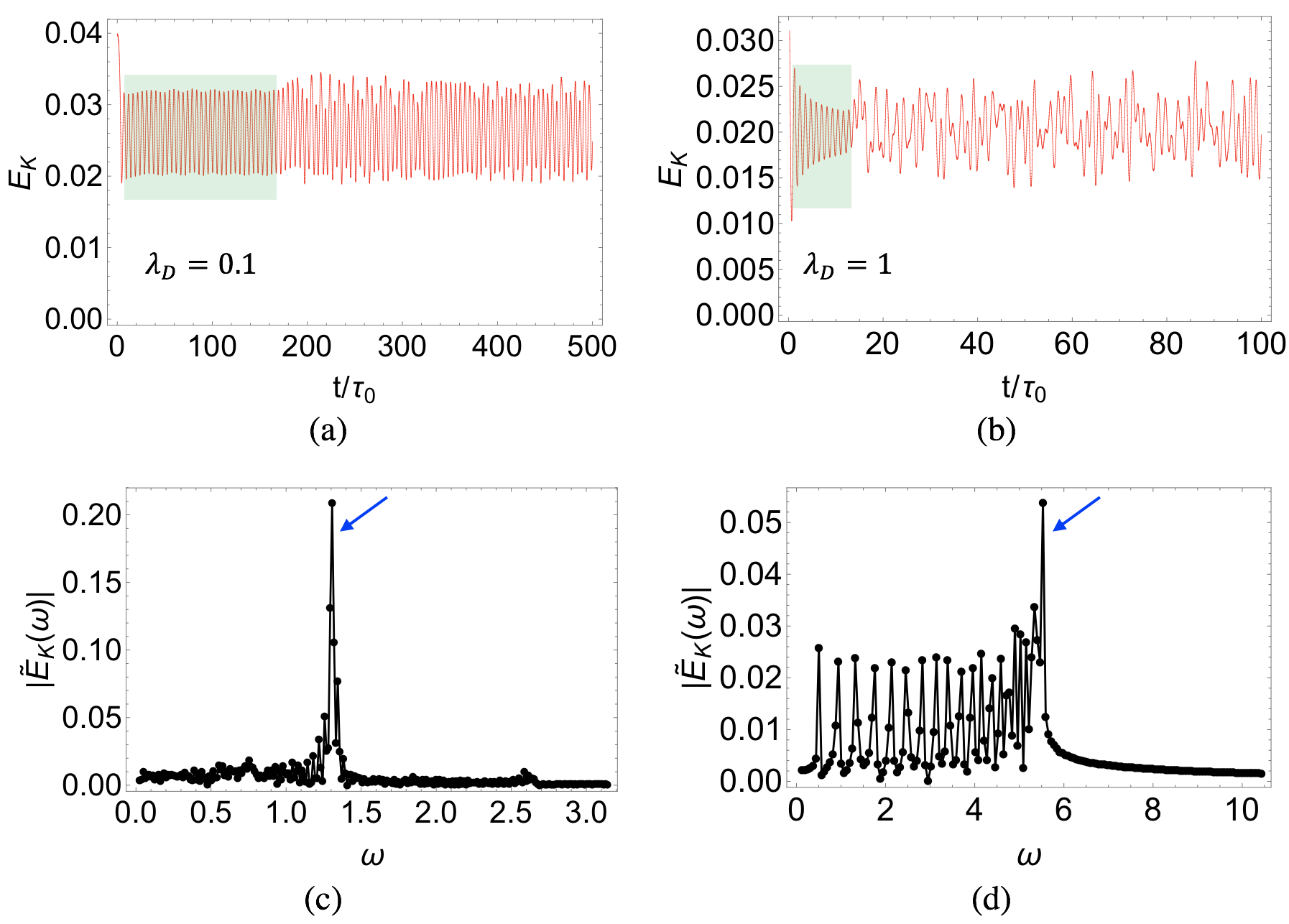}
  \caption{Spectral analysis of the kinetic energy of short- and long-range
  interacting systems.  The spectral analysis is based on 5000 recorded energy
  data. The spectra of the kinetic energy curves in (a) and (b) for
  $\lambda_D=0.1$ and $\lambda_D=1$ are presented in (c) and (d), respectively.
  More dynamical modes are excited in the latter system. The peak frequencies
  indicated by the arrows correspond to the initial regular oscillations in the
  green regions in (a) and (b). The energy is measured in the unit of
  $\epsilon_0=m(a_0/\tau_0)^2$. $\Gamma=0.1$.  $N=50$.   }
\label{energy}
\end{figure}

\section{Results and Discussion}

\subsection{Energy spectral analysis} We first track the motion of the disturbed
particle labelled as $i=1$. The recorded
trajectory $\theta_{i=1}(t)$ is presented in fig.~\ref{schematic}(b); the value
for $\theta_{i=1}(t)$ is scaled by the initial particle-particle angular
distance $\beta_0$ to show the relative displacement of the particles.
It clearly shows
that varying the range of interaction from $\lambda_D=0.1$ to $\lambda_D=1$
leads to distinct behaviors of the disturbed particle. The displacement curve (black) in
fig.~\ref{schematic}(b) for $\lambda_D=0.1$ exhibits stepped feature. 
Each fast forward motion with the abrupt increase of $\theta$ is
accompanied by a very slow backward motion. This featured rhythm in the
motion of the disturbed particle persists during the entire simulation time of
$\Delta t = 500\tau_0$. In the strong screening regime of small
$\lambda_D$, the interaction of adjacent particles resembles the two-body
collision in Newton's cradle, and it is thus expected that the steps in the
displacement curve of the disturbed particle would become flat. In contrast,
for $\lambda_D=1$, the motion of the disturbed particle is simultaneously
affected by several neighboring particles, resulting in the fine wave structure
in the trajectory as shown in the red curve in fig.~\ref{schematic}(b). To
exclude the possibility that the wave structure may be caused by any artifact
arising in simulations, we vary the initial speed, the number of particles and
the time step, and obtain consistent results.  $\theta_{i=1}(t)$-curves at
larger screening lengths are also featured with the fine wave structure. As
such, the single disturbed particle serves as a probe to reveal the subtle
dynamical effect associated with the long-range nature of the interaction.

Figures~\ref{energy}(a) and \ref{energy}(b) show the long-time kinetic energy
curves for $\lambda_D=0.1$ and $\lambda_D=1$. A common feature of both kinetic
energy curves is the sharply divided regular and irregular oscillations. The
regular regions are indicated by the light green boxes. It is observed that
the kinetic energy for $\lambda_D=1$ oscillates much faster than that for
$\lambda_D=0.1$. Furthermore, the amplitude of the oscillation is significantly
reduced in time for $\lambda_D=1$. In the region of irregular oscillation, the
kinetic energy curve for $\lambda_D=1$ is subject to stronger undulations.

To uncover the underlying dynamical modes, the kinetic energy curves in
figs.~\ref{energy}(a) and \ref{energy}(b) are fourier transformed, and the
results are presented in figs.~\ref{energy}(c) and \ref{energy}(d).  The peak
frequencies indicated by the arrows correspond to the initial regular
oscillations in figs.~\ref{energy}(a) and \ref{energy}(b), respectively. The
energy spectra reveal a significant difference between the short- and
long-range interactions: more dynamical modes are excited in the latter
system.  This phenomenon shall be attributed to the enhanced nonlinearity effect
under long-range interaction~\cite{scheck2010mechanics}.

\subsection{Mobilization process} To explore the physical origin of the dynamical
transition from the regular to the irregular oscillation of the kinetic energy
in fig.~\ref{energy}, we analyze the propagation of the disturbance on the
single particle. Through the persistent back-and-forth interactions with
neighboring particles, the initially disturbed particle mobilizes the remaining
particles in sequence. A key observation is that, prior to the complete
mobilization of the entire system, the variation of the total kinetic energy
follows the regular pattern in fig.~\ref{energy}.

\begin{figure}[t]  % h: put figure just in this position if possible
\centering 
\includegraphics[width=3.4in]{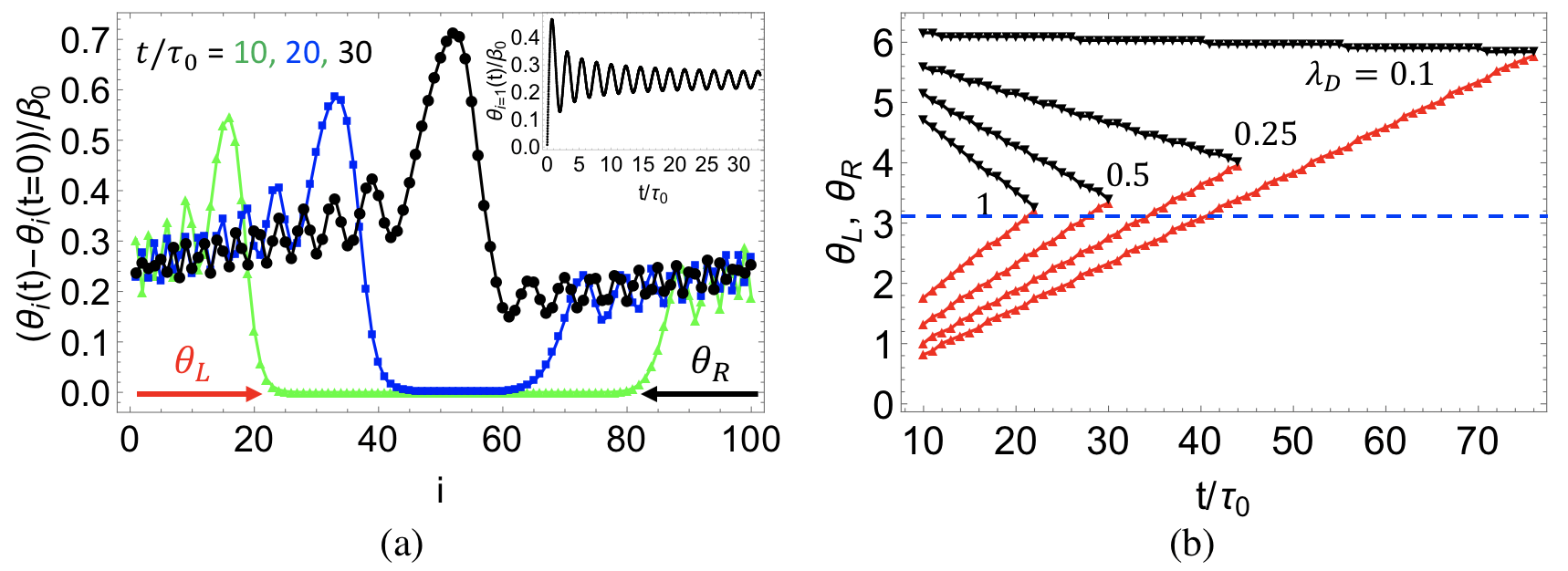}
  \caption{Mobilization of the initially static particles triggered by the
  disturbed particle. (a) Temporally varying $\theta_i(t)$ profiles, where $i$ is
  the label of particles in sequence around the ring. $\beta_0$ is
  particle-particle angular distance in the initial regular packing of the
  particles on the ring. The dynamics of the
  disturbed particle ($i=1$) is also presented in the inset. $\lambda_D=1$. (b)
  The mobilization process exhibits reflection symmetry with respect to the
  disturbed particle as the range of interaction increases. $\theta_L$ and
  $\theta_R$ represent the locations of the bilateral frontiers of the mobilized
  particles around the initially disturbed particle, as indicated by the arrows
  in (a). $N=100$.  $\Gamma=1$.  }
\label{mobilize}
\end{figure}

We further analyze the mobilization process quantitatively in terms of the two dynamical
variables $\theta_{{\textrm L}}$ and $\theta_{{\textrm R}}$ whose definitions
are shown in fig.~\ref{mobilize}(a). In fig.~\ref{mobilize}(a), $\theta_i(t)$ is
the position of particle $i$ at time $t$. $\theta_{{\textrm L}}$ and
$\theta_{{\textrm R}}$ are the locations of the bilateral frontiers of the
mobilized particles around the initially disturbed particle. Each curve in
fig.~\ref{mobilize}(a) represents an instantaneous particle configuration. The
trajectory of the initially disturbed particle is presented in the inset.
Comparison of the $\theta_i(t)$-curves at $t/\tau_0=10$ (green) and $t/\tau_0=20$
(blue) shows the movement of the frontiers of the mobilized particles as
characterized by the increase of $\theta_{{\textrm L}}$ and the decrease of
$\theta_{{\textrm R}}$.

The evolution of $\theta_{{\textrm L}}$ and $\theta_{{\textrm R}}$ at varying
$\lambda_D$ is plotted in fig.~\ref{mobilize}(b). All of the particles are
mobilized as the pair of the curves (red and black) meet at the ends. The dashed
horizontal line at $\theta=\pi$ indicates the reference position of the
diametrical particle with respect to the disturbed one.
Figure~\ref{mobilize}(b) shows the discrepancy of the bilateral mobilization
rates in the relatively short-range interacting systems. The particles
towards which the disturbed particle initially approaches are mobilized in a
much faster rate than those on the other side. To understand the origin of
the dynamical asymmetry, we notice that the loss of the momentum of the
disturbed particle in the first collision leads to a much milder second
collision with the other neighbor [see the curve of $\lambda=0.1$ in
fig.~\ref{schematic}(b)].  Here, for the long-range interacting
particles on the ring, the term collision refers to the approaching and
subsequent bouncing-away of two adjacent particles. The reflection symmetry
around the disturbed particle is thus broken. With the increase of $\lambda_D$,
the bilateral frontiers of the mobilized particles tend to advance at a common
rate as shown in fig.~\ref{mobilize}(b).  In other words, long-range interaction
preserves the reflection symmetry in the bilateral mobilization processes around
the disturbed particle. Furthermore, fig.~\ref{mobilize}(b) suggests
that the entire system would be mobilized in a faster fashion with the increase
of $\lambda_D$.

\begin{figure}[t]  % h: put figure just in this position if possible
\centering 
\includegraphics[width=3.2in]{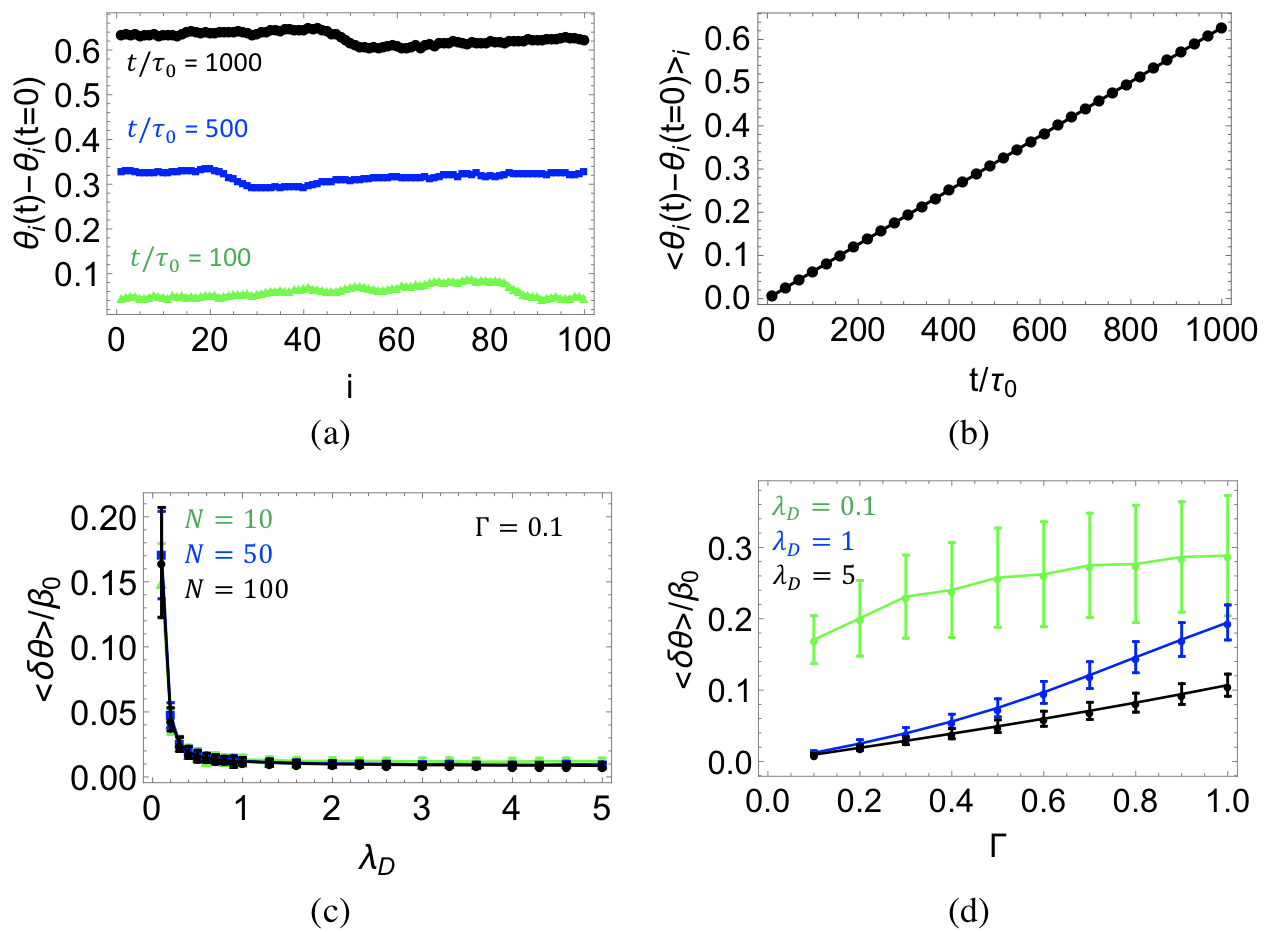}
  \caption{Suppression of the noise in the driven motion of particles under the
  long-range interaction. (a) The irregularity in the profile of the particle
  displacement $\theta_{i}(t)-\theta_{i}(t=0)$ indicates the
  presence of noise in the motion of particles. $\lambda_D=1$. $\Gamma=1$. (b)
  Plot of the mean particle displacement versus time. The averaging is over all
  the particles. The linear curve describes the uniform global rotation of the
  particles. (c) and (d) Dependence of the noise amplitude $\langle \delta
  \theta \rangle$ on the screening length $\lambda_D$ and the initial speed
  $\Gamma$ of the disturbed particle.  $\langle \delta \theta \rangle$ is the
  time-averaged standard deviation of the particle displacement over twenty
  particle configurations from $t\in[0, 200\tau_0]$. $\Gamma=0.1$ (c). $N=100$
  (d).  }
\label{dTheta}
\end{figure}

\begin{figure*}[t]  % h: put figure just in this position if possible
\centering 
\includegraphics[width=6.4in]{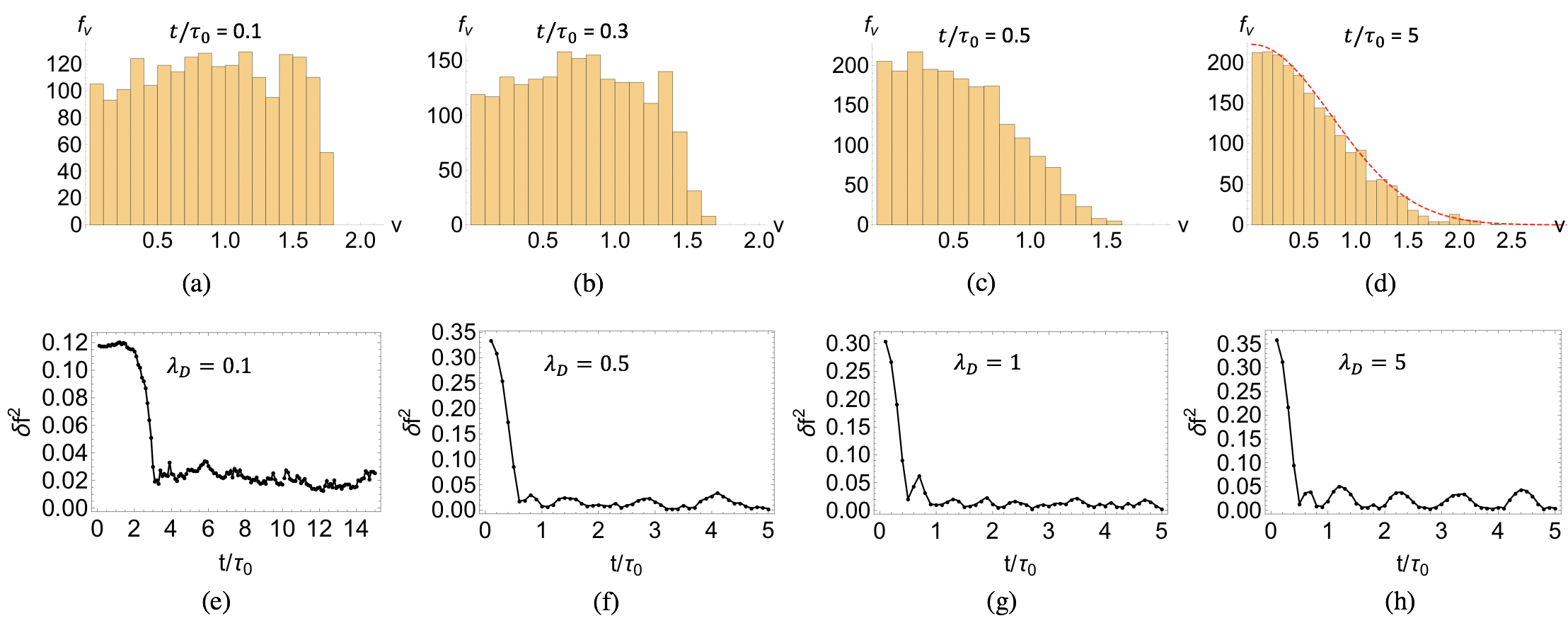}
  \caption{Relaxation process of the particle speed. (a)-(d)
  Typical instantaneous speed distributions at $\lambda_D=1$. (e)-(h) Plot of
  the $\delta f^2(t)$ curves at varying screening length. $\delta f^2$ is the
  deviation of an instantaneous speed distribution from the one-dimensional
  Maxwell-Boltzmann distribution. See the main text for more information.
  $N=2000$. $\Gamma_m=0.1$.  }
\label{fv}
\end{figure*}

\subsection{Global rotation and intrinsic noise} We further discuss the long-time
collective dynamics driven by the disturbed particle. Typical instantaneous
particle configurations are presented in fig.~\ref{dTheta}(a). The elevating
displacement curve implies the global rotation of the particles.
Note that the global rotation as characterized by the elevation of the
displacement curve occurs when all particles are mobilized
for $t/\tau_0 \gtrapprox 55$. Such a
dynamical state is characterized by the mean displacement $\langle
\theta_{i}(t)-\theta_{i}(t=0)\rangle_{i}$ in
fig.~\ref{dTheta}(b); the averaging is over all the particles.
Figure~\ref{dTheta}(b) shows that the mean displacement is a linear function of
time, indicating that the global rotation is uniform. The slope $\dot{\theta}_{rot}$ of the line in
fig.~\ref{dTheta}(b) is determined by the initial speed of the disturbed
particle. For the case in fig.~\ref{dTheta}, the relative difference between
$\dot{\theta}_{rot}$ and $v_{{\textrm {ini}}}/(R_0 N)$ is about $0.5\%$.
This small discrepancy can be attributed to the undulations on the $\theta_i(t)$
curves as shown in fig.~\ref{dTheta}(a). In the formation of the
uniform rotation, the initial kinetic energy input on the disturbed particle is
ultimately transferred to all the other particles. In this process, both the
energy and the angular momentum are conserved, as confirmed in
simulations. Note that in the limit of $N\rightarrow \infty$, the mean
kinetic energy on each particle becomes negligibly small. As such, the emergence of the
global rotation is a finite-size effect. The
motion of the particles for the duration of $\Delta t/\tau_0 = 1000$ ($\lambda_D=1$ and
$\Gamma=1$) is recorded in a movie (see the Supplemental Information).

The irregular undulations on the displacement curves in fig.~\ref{dTheta}(a),
which are referred to as noise in the following discussions, are
persistent during the entire simulation up to $t/\tau_0=1000$. Systematic
simulations at varying levels of the initial disturbance and with different
number of particles show that the presence of noise on the
$\theta_i(t)$-curve
is a common feature in the driven motion of particles. In other words, the
initial disturbance on the single particle cannot be fully transformed to a
perfectly uniform rotation of the system. Here, we remark that this phenomenon
bridges the dynamical and statistical aspects of the interacting particle
system. While the dynamics of the interacting particles is deterministic, the
noise arising on the displacement curve originates from the
randomization of particle speed under collisions. This is a spontaneous physical
process that is governed by the fundamental mechanical and statistical laws. As
such, the noise in the motion of particles is an intrinsic property of
interacting particle systems. Note that the noise may also be interpreted as the
superposition of dynamical modes, and the uniform rotation represents the
fundamental zero mode.

Is there any statistical regularity underlying the noise? Especially, how could
the range of interaction influence the statistical behavior of the noise? These
inquiries lead us to perform statistical analysis of the trajectories of motion.
Specifically, we compute the standard deviation $\delta \theta(t)$ of the
instantaneous particle configuration, and then take time average of
$\delta \theta(t)$, which is denoted as $\langle \delta \theta \rangle$. The
information of particle-particle interaction is encodes in the quantity of the
noise amplitude $\langle \delta \theta \rangle$. Analyzing $\langle \delta
\theta \rangle$ allows us to explore the connection of the range of interaction
and the random dynamics.

The dependence of $\langle \delta \theta \rangle$ on the screening length
$\lambda_D$ and the disturbance strength $\Gamma$ is shown in
fig.~\ref{dTheta}(c) and \ref{dTheta}(d).  From fig.~\ref{dTheta}(c), we see the
abrupt reduction of $\langle \delta \theta \rangle$ with the increase of the
screening length, indicating that the noise amplitude is significantly
suppressed under long-range interaction. The coincidence of the curves for the
different number of particles indicates that the amplitude of noise is
independent of the number of particles. The noise level is also suppressed under
weaker initial disturbance as shown in the plot of $\langle \delta \theta
\rangle$ versus $\Gamma$ in fig.~\ref{dTheta}(d). Extrapolation of the curves in
fig.~\ref{dTheta}(d) to the larger $\lambda_D$ regime suggests that the noise in
the displacement curve would be further reduced in the Coulomb interacting
systems.

\subsection{Dynamics of speed relaxation} In this section, we discuss the dynamical
process of speed redistribution among the particles. To
accelerate this process, all of the particles are disturbed. The initial speed of each particle is a uniform
random value in the range of $[-\Gamma_m, \Gamma_m]$. We work
in the rotating frame of reference where the total angular momentum is zero. It is found
that, regardless of the range of interaction, collisions of particles ultimately
lead to a stable speed distribution in the form of 
\begin{eqnarray}
  f_0(v) = A e^{-v^2/v_p^2},\label{f0}
\end{eqnarray} 
where $A$ is the normalization coefficient. Equation (\ref{f0}) is essentially
the one-dimensional Maxwell-Boltzmann distribution. In
fig.~\ref{fv}(a)-\ref{fv}(d), we show the typical instantaneous speed
distributions in the relaxation process for $\lambda_D=1$. The red dashed
fitting curve in fig.~\ref{fv}(d) has the functional form of eq.(\ref{f0}). For
the case of the single-particle driven motion, the speed distribution also
evolves towards the equilibrium distribution as in eq.(\ref{f0}); the relaxation
time for the system of $\lambda_D=1$ and $\Gamma=1$ is several hundred times of
the characteristic time $\tau_0$. Here, it is of interest to note that, as a
statistical law, collisions could lead to the concentration of kinetic energy on
a small number of particles with high speed in the equilibrium state; any
particle in the ring system has the chance of becoming the high-speed one.

To quantitatively characterize the dynamics of speed relaxation, we propose 
the quantity $\delta f^2$ defined as:
\begin{eqnarray}
  \delta f^2 = \frac{\sum_{v_i} (f(v_i) - f_{0}(v_i))^2}{\sum_{v_i}
  f_{0}(v_i)^2},
\end{eqnarray}
where $f(v)$ is the numerically obtained instantaneous speed distribution during
the relaxation process, and $f_0(v)$ is the equilibrium speed distribution.
Figures~\ref{fv}(e)-\ref{fv}(h) show the temporal variation of $\delta f^2$ at
varying screening length. We see that the relaxation time significantly
reduces from about $3\tau_0$ [see fig.~\ref{fv}(e)] to a fraction of $\tau_0$
[see figs.~\ref{fv}(f)-~\ref{fv}(h)] with the increase of $\lambda_D$. Besides
the fast relaxation process, the fluctuation of the $\delta f^2(t)$ curve seems
exhibiting some degree of periodicity in the long-range interacting system of
$\lambda_D=5$ in fig.~\ref{fv}(h), implying the presence of some ordered
dynamical structure.

\section{Conclusion}

In summary, we investigated the dynamical response of the system to the
disturbance imposed on a single particle, and reveal the featured dynamics
brought by the long-range interaction. Specifically, we highlight the efficient transformation
of the disturbance into the uniform global rotation of the system, the
suppression of the intrinsic noise, and the fast relaxation of particle speed
under the long-range interaction. These results advance our understanding on the
dynamical effects of long-range interaction. The simple 1D ring model may be
employed to address the fundamental questions on the interface of microscopic
dynamics and statistical physics of long-range interacting systems.

\acknowledgments

This work was supported by the National Natural Science Foundation of China
(Grants No. BC4190050).  The author acknowledges the support from the Student
Innovation Center at Shanghai Jiao Tong University.

%\bibliography{/Users/zyao/Documents/Researches/ref_jab_5}

\begin{thebibliography}{42}
\expandafter\ifx\csname natexlab\endcsname\relax\def\natexlab#1{#1}\fi
\expandafter\ifx\csname bibnamefont\endcsname\relax
  \def\bibnamefont#1{#1}\fi
\expandafter\ifx\csname bibfnamefont\endcsname\relax
  \def\bibfnamefont#1{#1}\fi
\expandafter\ifx\csname citenamefont\endcsname\relax
  \def\citenamefont#1{#1}\fi
\expandafter\ifx\csname url\endcsname\relax
  \def\url#1{\texttt{#1}}\fi
\expandafter\ifx\csname urlprefix\endcsname\relax\def\urlprefix{URL }\fi
\providecommand{\bibinfo}[2]{#2}
\providecommand{\eprint}[2][]{\url{#2}}

\bibitem[{\citenamefont{Lynden-Bell et~al.}(1968)\citenamefont{Lynden-Bell,
  Wood, and Royal}}]{lynden1968gravo}
\bibinfo{author}{\bibfnamefont{D.}~\bibnamefont{Lynden-Bell}},
  \bibinfo{author}{\bibfnamefont{R.}~\bibnamefont{Wood}}, \bibnamefont{and}
  \bibinfo{author}{\bibfnamefont{A.}~\bibnamefont{Royal}},
  \bibinfo{journal}{Monthly Notices of the Royal Astronomical Society}
  \textbf{\bibinfo{volume}{138}}, \bibinfo{pages}{495} (\bibinfo{year}{1968}).

\bibitem[{\citenamefont{Padmanabhan}(1990)}]{padmanabhan1990statistical}
\bibinfo{author}{\bibfnamefont{T.}~\bibnamefont{Padmanabhan}},
  \bibinfo{journal}{Phys. Rep.} \textbf{\bibinfo{volume}{188}},
  \bibinfo{pages}{285} (\bibinfo{year}{1990}).

\bibitem[{\citenamefont{Joyce and
  Worrakitpoonpon}(2011)}]{joyce2011quasistationary}
\bibinfo{author}{\bibfnamefont{M.}~\bibnamefont{Joyce}} \bibnamefont{and}
  \bibinfo{author}{\bibfnamefont{T.}~\bibnamefont{Worrakitpoonpon}},
  \bibinfo{journal}{Physical Review E} \textbf{\bibinfo{volume}{84}},
  \bibinfo{pages}{011139} (\bibinfo{year}{2011}).

\bibitem[{\citenamefont{Lighthill}(1976)}]{lighthill1976flagellar}
\bibinfo{author}{\bibfnamefont{J.}~\bibnamefont{Lighthill}},
  \bibinfo{journal}{SIAM Rev} \textbf{\bibinfo{volume}{18}},
  \bibinfo{pages}{161} (\bibinfo{year}{1976}).

\bibitem[{\citenamefont{Chattopadhyay and Wu}(2009)}]{chattopadhyay2009effect}
\bibinfo{author}{\bibfnamefont{S.}~\bibnamefont{Chattopadhyay}}
  \bibnamefont{and} \bibinfo{author}{\bibfnamefont{X.-L.} \bibnamefont{Wu}},
  \bibinfo{journal}{Biophys. J.} \textbf{\bibinfo{volume}{96}},
  \bibinfo{pages}{2023} (\bibinfo{year}{2009}).

\bibitem[{\citenamefont{Dallaston et~al.}(2018)\citenamefont{Dallaston,
  Fontelos, Tseluiko, and Kalliadasis}}]{dallaston2018discrete}
\bibinfo{author}{\bibfnamefont{M.~C.} \bibnamefont{Dallaston}},
  \bibinfo{author}{\bibfnamefont{M.~A.} \bibnamefont{Fontelos}},
  \bibinfo{author}{\bibfnamefont{D.}~\bibnamefont{Tseluiko}}, \bibnamefont{and}
  \bibinfo{author}{\bibfnamefont{S.}~\bibnamefont{Kalliadasis}},
  \bibinfo{journal}{Phys. Rev. Lett.} \textbf{\bibinfo{volume}{120}},
  \bibinfo{pages}{034505} (\bibinfo{year}{2018}).

\bibitem[{\citenamefont{Walker et~al.}(2011)\citenamefont{Walker, Kowalczyk,
  Olvera de~la Cruz, and Grzybowski}}]{Walker2011}
\bibinfo{author}{\bibfnamefont{D.~A.} \bibnamefont{Walker}},
  \bibinfo{author}{\bibfnamefont{B.}~\bibnamefont{Kowalczyk}},
  \bibinfo{author}{\bibfnamefont{M.}~\bibnamefont{Olvera de~la Cruz}},
  \bibnamefont{and} \bibinfo{author}{\bibfnamefont{B.~A.}
  \bibnamefont{Grzybowski}}, \bibinfo{journal}{Nanoscale}
  \textbf{\bibinfo{volume}{3}}, \bibinfo{pages}{1316} (\bibinfo{year}{2011}).

\bibitem[{\citenamefont{Christodoulidi
  et~al.}(2014)\citenamefont{Christodoulidi, Tsallis, and
  Bountis}}]{christodoulidi2014fermi}
\bibinfo{author}{\bibfnamefont{H.}~\bibnamefont{Christodoulidi}},
  \bibinfo{author}{\bibfnamefont{C.}~\bibnamefont{Tsallis}}, \bibnamefont{and}
  \bibinfo{author}{\bibfnamefont{T.}~\bibnamefont{Bountis}},
  \bibinfo{journal}{Europhys. Lett.} \textbf{\bibinfo{volume}{108}},
  \bibinfo{pages}{40006} (\bibinfo{year}{2014}).

\bibitem[{\citenamefont{Juh{\'a}sz et~al.}(2014)\citenamefont{Juh{\'a}sz,
  Kov{\'a}cs, and Igl{\'o}i}}]{juhasz2014random}
\bibinfo{author}{\bibfnamefont{R.}~\bibnamefont{Juh{\'a}sz}},
  \bibinfo{author}{\bibfnamefont{I.~A.} \bibnamefont{Kov{\'a}cs}},
  \bibnamefont{and}
  \bibinfo{author}{\bibfnamefont{F.}~\bibnamefont{Igl{\'o}i}},
  \bibinfo{journal}{Europhys. Lett.} \textbf{\bibinfo{volume}{107}},
  \bibinfo{pages}{47008} (\bibinfo{year}{2014}).

\bibitem[{\citenamefont{Jadhao et~al.}(2015)\citenamefont{Jadhao, Yao, Thomas,
  and Olvera de~la Cruz}}]{jadhao2015coulomb}
\bibinfo{author}{\bibfnamefont{V.}~\bibnamefont{Jadhao}},
  \bibinfo{author}{\bibfnamefont{Z.}~\bibnamefont{Yao}},
  \bibinfo{author}{\bibfnamefont{C.~K.} \bibnamefont{Thomas}},
  \bibnamefont{and} \bibinfo{author}{\bibfnamefont{M.}~\bibnamefont{Olvera
  de~la Cruz}}, \bibinfo{journal}{Phys. Rev. E} \textbf{\bibinfo{volume}{91}},
  \bibinfo{pages}{032305} (\bibinfo{year}{2015}).

\bibitem[{\citenamefont{Mauri and Katsnelson}(2019)}]{mauri2019thermal}
\bibinfo{author}{\bibfnamefont{A.}~\bibnamefont{Mauri}} \bibnamefont{and}
  \bibinfo{author}{\bibfnamefont{M.~I.} \bibnamefont{Katsnelson}},
  \bibinfo{journal}{Ann. Phys.} \textbf{\bibinfo{volume}{412}},
  \bibinfo{pages}{168016} (\bibinfo{year}{2019}).

\bibitem[{\citenamefont{Yao}(2019)}]{yao2019command}
\bibinfo{author}{\bibfnamefont{Z.}~\bibnamefont{Yao}}, \bibinfo{journal}{Phys.
  Rev. Lett.} \textbf{\bibinfo{volume}{122}}, \bibinfo{pages}{228002}
  (\bibinfo{year}{2019}).

\bibitem[{\citenamefont{Holm et~al.}(2001)\citenamefont{Holm, K{\'e}kicheff,
  and Podgornik}}]{Holm2001}
\bibinfo{author}{\bibfnamefont{C.}~\bibnamefont{Holm}},
  \bibinfo{author}{\bibfnamefont{P.}~\bibnamefont{K{\'e}kicheff}},
  \bibnamefont{and}
  \bibinfo{author}{\bibfnamefont{R.}~\bibnamefont{Podgornik}},
  \emph{\bibinfo{title}{Electrostatic Effects in Soft Matter and Biophysics}}
  (\bibinfo{publisher}{Springer, Berlin}, \bibinfo{year}{2001}).

\bibitem[{\citenamefont{Levin}(2002)}]{Levin2002}
\bibinfo{author}{\bibfnamefont{Y.}~\bibnamefont{Levin}}, \bibinfo{journal}{Rep.
  Prog. Phys.} \textbf{\bibinfo{volume}{65}}, \bibinfo{pages}{1577}
  (\bibinfo{year}{2002}).

\bibitem[{\citenamefont{Messina}(2008)}]{messina2008electrostatics}
\bibinfo{author}{\bibfnamefont{R.}~\bibnamefont{Messina}}, \bibinfo{journal}{J.
  Phys.: Condens. Matter} \textbf{\bibinfo{volume}{21}},
  \bibinfo{pages}{113102} (\bibinfo{year}{2008}).

\bibitem[{\citenamefont{Xing}(2011)}]{xing2011poisson}
\bibinfo{author}{\bibfnamefont{X.}~\bibnamefont{Xing}},
  \bibinfo{journal}{Physical Review E} \textbf{\bibinfo{volume}{83}},
  \bibinfo{pages}{041410} (\bibinfo{year}{2011}).

\bibitem[{\citenamefont{Yao and Olvera de~la
  Cruz}(2016)}]{yao2016electrostatics}
\bibinfo{author}{\bibfnamefont{Z.}~\bibnamefont{Yao}} \bibnamefont{and}
  \bibinfo{author}{\bibfnamefont{M.}~\bibnamefont{Olvera de~la Cruz}},
  \bibinfo{journal}{Phys. Rev. Lett.} \textbf{\bibinfo{volume}{116}},
  \bibinfo{pages}{148101} (\bibinfo{year}{2016}).

\bibitem[{\citenamefont{Gao et~al.}(2019)\citenamefont{Gao, Kewalramani,
  Valencia, Li, McCourt, Olvera de~la Cruz, and Bedzyk}}]{gao2019electrostatic}
\bibinfo{author}{\bibfnamefont{C.}~\bibnamefont{Gao}},
  \bibinfo{author}{\bibfnamefont{S.}~\bibnamefont{Kewalramani}},
  \bibinfo{author}{\bibfnamefont{D.~M.} \bibnamefont{Valencia}},
  \bibinfo{author}{\bibfnamefont{H.}~\bibnamefont{Li}},
  \bibinfo{author}{\bibfnamefont{J.~M.} \bibnamefont{McCourt}},
  \bibinfo{author}{\bibfnamefont{M.}~\bibnamefont{Olvera de~la Cruz}},
  \bibnamefont{and} \bibinfo{author}{\bibfnamefont{M.~J.}
  \bibnamefont{Bedzyk}}, \bibinfo{journal}{Proceedings of the National Academy
  of Sciences} \textbf{\bibinfo{volume}{116}}, \bibinfo{pages}{22030}
  (\bibinfo{year}{2019}).

\bibitem[{\citenamefont{Bouchet and Barre}(2005)}]{bouchet2005classification}
\bibinfo{author}{\bibfnamefont{F.}~\bibnamefont{Bouchet}} \bibnamefont{and}
  \bibinfo{author}{\bibfnamefont{J.}~\bibnamefont{Barre}}, \bibinfo{journal}{J.
  Stat. Phys.} \textbf{\bibinfo{volume}{118}}, \bibinfo{pages}{1073}
  (\bibinfo{year}{2005}).

\bibitem[{\citenamefont{Levin}(2005)}]{levin2005strange}
\bibinfo{author}{\bibfnamefont{Y.}~\bibnamefont{Levin}},
  \bibinfo{journal}{Physica A: Statistical Mechanics and its Applications}
  \textbf{\bibinfo{volume}{352}}, \bibinfo{pages}{43} (\bibinfo{year}{2005}).

\bibitem[{\citenamefont{Rocha~Filho et~al.}(2005)\citenamefont{Rocha~Filho,
  Figueiredo, and Amato}}]{rocha2005entropy}
\bibinfo{author}{\bibfnamefont{T.}~\bibnamefont{Rocha~Filho}},
  \bibinfo{author}{\bibfnamefont{A.}~\bibnamefont{Figueiredo}},
  \bibnamefont{and} \bibinfo{author}{\bibfnamefont{M.~A.} \bibnamefont{Amato}},
  \bibinfo{journal}{Phys. Rev. Lett.} \textbf{\bibinfo{volume}{95}},
  \bibinfo{pages}{190601} (\bibinfo{year}{2005}).

\bibitem[{\citenamefont{Pluchino et~al.}(2007)\citenamefont{Pluchino,
  Rapisarda, and Tsallis}}]{pluchino2007nonergodicity}
\bibinfo{author}{\bibfnamefont{A.}~\bibnamefont{Pluchino}},
  \bibinfo{author}{\bibfnamefont{A.}~\bibnamefont{Rapisarda}},
  \bibnamefont{and} \bibinfo{author}{\bibfnamefont{C.}~\bibnamefont{Tsallis}},
  \bibinfo{journal}{Europhys. Lett.} \textbf{\bibinfo{volume}{80}},
  \bibinfo{pages}{26002} (\bibinfo{year}{2007}).

\bibitem[{\citenamefont{Joyce et~al.}(1971)\citenamefont{Joyce, Knorr, and
  Meier}}]{joyce1971numerical}
\bibinfo{author}{\bibfnamefont{G.}~\bibnamefont{Joyce}},
  \bibinfo{author}{\bibfnamefont{G.}~\bibnamefont{Knorr}}, \bibnamefont{and}
  \bibinfo{author}{\bibfnamefont{H.~K.} \bibnamefont{Meier}},
  \bibinfo{journal}{J. Comput. Phys.} \textbf{\bibinfo{volume}{8}},
  \bibinfo{pages}{53} (\bibinfo{year}{1971}).

\bibitem[{\citenamefont{Pakter and Levin}(2017)}]{pakter2017entropy}
\bibinfo{author}{\bibfnamefont{R.}~\bibnamefont{Pakter}} \bibnamefont{and}
  \bibinfo{author}{\bibfnamefont{Y.}~\bibnamefont{Levin}}, \bibinfo{journal}{J.
  Stat. Mech: Theory Exp.} \textbf{\bibinfo{volume}{2017}},
  \bibinfo{pages}{044001} (\bibinfo{year}{2017}).

\bibitem[{\citenamefont{Cirto et~al.}(2018)\citenamefont{Cirto, Rodr{\'\i}guez,
  Nobre, and Tsallis}}]{cirto2018validity}
\bibinfo{author}{\bibfnamefont{L.~J.} \bibnamefont{Cirto}},
  \bibinfo{author}{\bibfnamefont{A.}~\bibnamefont{Rodr{\'\i}guez}},
  \bibinfo{author}{\bibfnamefont{F.~D.} \bibnamefont{Nobre}}, \bibnamefont{and}
  \bibinfo{author}{\bibfnamefont{C.}~\bibnamefont{Tsallis}},
  \bibinfo{journal}{Europhys. Lett.} \textbf{\bibinfo{volume}{123}},
  \bibinfo{pages}{30003} (\bibinfo{year}{2018}).

\bibitem[{\citenamefont{Toner and Tu}(1995)}]{toner1995long}
\bibinfo{author}{\bibfnamefont{J.}~\bibnamefont{Toner}} \bibnamefont{and}
  \bibinfo{author}{\bibfnamefont{Y.}~\bibnamefont{Tu}}, \bibinfo{journal}{Phys.
  Rev. Lett.} \textbf{\bibinfo{volume}{75}}, \bibinfo{pages}{4326}
  (\bibinfo{year}{1995}).

\bibitem[{\citenamefont{Zhang et~al.}(2013)\citenamefont{Zhang, Jha, and Olvera
  de~la Cruz}}]{zhang2013non}
\bibinfo{author}{\bibfnamefont{R.}~\bibnamefont{Zhang}},
  \bibinfo{author}{\bibfnamefont{P.}~\bibnamefont{Jha}}, \bibnamefont{and}
  \bibinfo{author}{\bibfnamefont{M.}~\bibnamefont{Olvera de~la Cruz}},
  \bibinfo{journal}{Soft Matter} \textbf{\bibinfo{volume}{9}},
  \bibinfo{pages}{5042} (\bibinfo{year}{2013}).

\bibitem[{\citenamefont{Grzybowski et~al.}(2009)\citenamefont{Grzybowski,
  Wilmer, Kim, Browne, and Bishop}}]{grzybowski2009self}
\bibinfo{author}{\bibfnamefont{B.~A.} \bibnamefont{Grzybowski}},
  \bibinfo{author}{\bibfnamefont{C.~E.} \bibnamefont{Wilmer}},
  \bibinfo{author}{\bibfnamefont{J.}~\bibnamefont{Kim}},
  \bibinfo{author}{\bibfnamefont{K.~P.} \bibnamefont{Browne}},
  \bibnamefont{and} \bibinfo{author}{\bibfnamefont{K.~J.}
  \bibnamefont{Bishop}}, \bibinfo{journal}{Soft Matter}
  \textbf{\bibinfo{volume}{5}}, \bibinfo{pages}{1110} (\bibinfo{year}{2009}).

\bibitem[{\citenamefont{Kadanoff}(1999)}]{Kadanoff1999}
\bibinfo{author}{\bibfnamefont{L.~P.} \bibnamefont{Kadanoff}},
  \emph{\bibinfo{title}{From Order to Chaos II}} (\bibinfo{publisher}{World
  Scientific}, \bibinfo{year}{1999}).

\bibitem[{\citenamefont{Chen}(2012)}]{chen2012introduction}
\bibinfo{author}{\bibfnamefont{F.~F.} \bibnamefont{Chen}},
  \emph{\bibinfo{title}{Introduction to plasma physics}}
  (\bibinfo{publisher}{Springer Science \& Business Media},
  \bibinfo{year}{2012}).

\bibitem[{\citenamefont{R{\"u}tzel et~al.}(2003)\citenamefont{R{\"u}tzel, Lee,
  and Raman}}]{rutzel2003nonlinear}
\bibinfo{author}{\bibfnamefont{S.}~\bibnamefont{R{\"u}tzel}},
  \bibinfo{author}{\bibfnamefont{S.~I.} \bibnamefont{Lee}}, \bibnamefont{and}
  \bibinfo{author}{\bibfnamefont{A.}~\bibnamefont{Raman}},
  \bibinfo{journal}{Proc. R. Soc. London, Ser. A}
  \textbf{\bibinfo{volume}{459}}, \bibinfo{pages}{1925} (\bibinfo{year}{2003}).

\bibitem[{\citenamefont{Campa et~al.}(2014)\citenamefont{Campa, Dauxois,
  Fanelli, and Ruffo}}]{campa2014physics}
\bibinfo{author}{\bibfnamefont{A.}~\bibnamefont{Campa}},
  \bibinfo{author}{\bibfnamefont{T.}~\bibnamefont{Dauxois}},
  \bibinfo{author}{\bibfnamefont{D.}~\bibnamefont{Fanelli}}, \bibnamefont{and}
  \bibinfo{author}{\bibfnamefont{S.}~\bibnamefont{Ruffo}},
  \emph{\bibinfo{title}{Physics of Long-Range Interacting Systems}}
  (\bibinfo{publisher}{Oxford University Press, Oxford, UK},
  \bibinfo{year}{2014}).

\bibitem[{\citenamefont{Boltzmann}(1964)}]{boltzmann1964lectures}
\bibinfo{author}{\bibfnamefont{L.}~\bibnamefont{Boltzmann}},
  \emph{\bibinfo{title}{Lectures On Gas Theory}}
  (\bibinfo{publisher}{University of California Press, Berkeley},
  \bibinfo{year}{1964}).

\bibitem[{\citenamefont{Rapaport}(2004)}]{rapaport2004art}
\bibinfo{author}{\bibfnamefont{D.}~\bibnamefont{Rapaport}},
  \emph{\bibinfo{title}{The Art of Molecular Dynamics Simulation}}
  (\bibinfo{publisher}{Cambridge University Press, Cambridge, UK},
  \bibinfo{year}{2004}).

\bibitem[{\citenamefont{Feix and Bertrand}(2005)}]{feix2005universal}
\bibinfo{author}{\bibfnamefont{M.}~\bibnamefont{Feix}} \bibnamefont{and}
  \bibinfo{author}{\bibfnamefont{P.}~\bibnamefont{Bertrand}},
  \bibinfo{journal}{Transp. Theory Stat. Phys.} \textbf{\bibinfo{volume}{34}},
  \bibinfo{pages}{7} (\bibinfo{year}{2005}).

\bibitem[{\citenamefont{Sota et~al.}(2001)\citenamefont{Sota, Iguchi, Morikawa,
  Tatekawa, and Maeda}}]{sota2001origin}
\bibinfo{author}{\bibfnamefont{Y.}~\bibnamefont{Sota}},
  \bibinfo{author}{\bibfnamefont{O.}~\bibnamefont{Iguchi}},
  \bibinfo{author}{\bibfnamefont{M.}~\bibnamefont{Morikawa}},
  \bibinfo{author}{\bibfnamefont{T.}~\bibnamefont{Tatekawa}}, \bibnamefont{and}
  \bibinfo{author}{\bibfnamefont{K.-i.} \bibnamefont{Maeda}},
  \bibinfo{journal}{Phys. Rev. E} \textbf{\bibinfo{volume}{64}},
  \bibinfo{pages}{056133} (\bibinfo{year}{2001}).

\bibitem[{\citenamefont{Sokolov et~al.}(2011)\citenamefont{Sokolov, Frydel,
  Grier, Diamant, and Roichman}}]{sokolov2011hydrodynamic}
\bibinfo{author}{\bibfnamefont{Y.}~\bibnamefont{Sokolov}},
  \bibinfo{author}{\bibfnamefont{D.}~\bibnamefont{Frydel}},
  \bibinfo{author}{\bibfnamefont{D.~G.} \bibnamefont{Grier}},
  \bibinfo{author}{\bibfnamefont{H.}~\bibnamefont{Diamant}}, \bibnamefont{and}
  \bibinfo{author}{\bibfnamefont{Y.}~\bibnamefont{Roichman}},
  \bibinfo{journal}{Phys. Rev. Lett.} \textbf{\bibinfo{volume}{107}},
  \bibinfo{pages}{158302} (\bibinfo{year}{2011}).

\bibitem[{\citenamefont{Nagar and Roichman}(2014)}]{nagar2014collective}
\bibinfo{author}{\bibfnamefont{H.}~\bibnamefont{Nagar}} \bibnamefont{and}
  \bibinfo{author}{\bibfnamefont{Y.}~\bibnamefont{Roichman}},
  \bibinfo{journal}{Phys. Rev. E} \textbf{\bibinfo{volume}{90}},
  \bibinfo{pages}{042302} (\bibinfo{year}{2014}).

\bibitem[{\citenamefont{Williams et~al.}(2016)\citenamefont{Williams,
  O{\u{g}}uz, Speck, Bartlett, L{\"o}wen, and
  Royall}}]{williams2016transmission}
\bibinfo{author}{\bibfnamefont{I.}~\bibnamefont{Williams}},
  \bibinfo{author}{\bibfnamefont{E.~C.} \bibnamefont{O{\u{g}}uz}},
  \bibinfo{author}{\bibfnamefont{T.}~\bibnamefont{Speck}},
  \bibinfo{author}{\bibfnamefont{P.}~\bibnamefont{Bartlett}},
  \bibinfo{author}{\bibfnamefont{H.}~\bibnamefont{L{\"o}wen}},
  \bibnamefont{and} \bibinfo{author}{\bibfnamefont{C.~P.}
  \bibnamefont{Royall}}, \bibinfo{journal}{Nat. Phys.}
  \textbf{\bibinfo{volume}{12}}, \bibinfo{pages}{98} (\bibinfo{year}{2016}).

\bibitem[{\citenamefont{Debye}(1923)}]{debye1923theory}
\bibinfo{author}{\bibfnamefont{P.}~\bibnamefont{Debye}},
  \bibinfo{journal}{Physikalische Zeitschrift} \textbf{\bibinfo{volume}{24}},
  \bibinfo{pages}{185} (\bibinfo{year}{1923}).

\bibitem[{\citenamefont{Dobrynin and Rubinstein}(2005)}]{Dobrynin2005}
\bibinfo{author}{\bibfnamefont{A.}~\bibnamefont{Dobrynin}} \bibnamefont{and}
  \bibinfo{author}{\bibfnamefont{M.}~\bibnamefont{Rubinstein}},
  \bibinfo{journal}{Prog. Polym. Sci.} \textbf{\bibinfo{volume}{30}},
  \bibinfo{pages}{1049} (\bibinfo{year}{2005}).

\bibitem[{\citenamefont{Scheck}(2010)}]{scheck2010mechanics}
\bibinfo{author}{\bibfnamefont{F.}~\bibnamefont{Scheck}},
  \emph{\bibinfo{title}{Mechanics: from Newton's Laws to Deterministic Chaos}}
  (\bibinfo{publisher}{Springer Science \& Business Media},
  \bibinfo{year}{2010}).

\end{thebibliography}
\end{document}